**Free Volume cannot Explain the Spatial Heterogeneity of Debye-Waller factors**

**in a Glass-Forming Binary Alloy.**

Asaph Widmer-Cooper and Peter Harrowell

School of Chemistry, University of Sydney, New South Wales, 2006, Australia

**Abstract**

We examine the relation between the free volume per particle and the variance of the particle position, equivalent to a local Debye-Waller (DW) factor for a 2D glass-forming alloy using molecular dynamics simulations. We find that the latter quantity exhibits significant spatial heterogeneity despite involving trajectories two orders of magnitude shorter than those typically used to measure such heterogeneities. We find that the free volume exhibits no significant spatial correlation with the local DW factor. We conclude that the spatial variation in local free volume is not the cause of the short time dynamic heterogeneity.

**1. Introduction**

Some aspect of the structure in each glass-forming alloy determines the observed slow particle dynamics. For example, strong correlations are observed between the increase in shear viscosity and large angle scattering structure in metallic alloys following a temperature quench [1]. The most widely used expression of this 'aspect of the structure' is the free volume [2]. (The concept of "shear transition zones"[3] is also used in the context of non-equilibrium mechanical response.)  The generation or disappearance of free volume has been invoked to explain shear banding [4] and positron annihilation [5] in metallic glasses.

Despite its popularity, there remains a persistent problem concerning the application of free volume to describe dynamics. Given the success of the free volume concept at a phenomenological level, we can pose this problem as follows: What is the relationship between the geometric free volume (a quantity that can be well defined at the atomic scale) and the phenomenological free volume (a macroscopic quantity that is derived from the bulk density)?  This question is the focus of this paper.

Phenomenological correlations are the staple of the glass field. A correlation between two variables, however, does not prove a causal link. We are interested in establishing whether or not geometrical free volume can explain the spatial variation in particle mobility in a dense amorphous alloy. To do this we propose that for a causal link to exist it is necessary that a strong *microscopic* correlation exist between free volume and dynamics.

Here we shall characterise the dynamics by particle displacements over a short time, one corresponding to the middle of the plateau region in the mean-squared displacement and in the incoherent scattering function.  A standard measure of displacement fluctuations in solids is the Debye-Waller (DW) factor, defined as the mean squared deviation of an atom from its equilibrium position averaged over all particles. Thus, one may write the DW factor as $<< (< \vec{r}_i > - \vec{r}_i(t))^2 >>$, where $\vec{r}_i$ is the instantaneous position of particle $i$, the inner angle brackets $<>$ and $< \vec{r}_i >$ refer to the time average and the outer brackets denote an average over particles. Note that the DW factor is the variance of the particle position rather than the mean squared displacement from an initial position. To calculate a DW factor for each individual particle in a configuration, we use a similar definition, except now the outer brackets denote an average over an isoconfigurational ensemble of runs. To generate an isoconfigurational ensemble, we shall perform 100 runs from the same initial configuration but with different initial momenta, randomly assigned from the correct Maxwell-Boltzmann distribution [6]. We shall refer to the variance of the position of a single particle for the distribution of positions generated by these multiple runs as the particle Debye-Waller (DW) factor, i.e.

$$DW_i = << (\vec{r}_i(t) - <\vec{r}_i>_{time})^2 >_{time} >_{isoconfig} \qquad (1)$$

where the time average is taken over the selected time interval and the isoconfigurational average is taken over the multiple trajectories. There is experimental evidence that the short time dynamics of particles can provide information about the long time behaviour of the system. Buchenau and Zorn [7] have reported that in selenium the mean squared particle displacement $<u^2>$ scales with the viscosity as

$$\eta = \eta_o \exp(C/[<u^2>_{cryt} - <u^2>_{liquid}])$$

over many orders of magnitude in η. Subsequently, a range of polymeric and small molecule glass formers [8] have also exhibited strong correlations between the short time fluctuations associated with the Debye-Waller factor and the viscosity. The significance of these results with regards the subject of this paper is that it suggests that the particle Debye-Waller factor we introduce here is not only a measure of local dynamics but may provide useful information about the long time relaxation of the amorphous system. We shall explore this idea elsewhere.

Recently, Starr et. al. [9] found a power law relation between average free volume and the bulk averaged short time mean squared displacement for monomers in a "bead-spring" model of a glass-forming polymer over a range of temperatures. This success of the free volume idea was qualified by their failure to find any significant correlation between the local free volume of a specific monomer and its mobility. We shall now consider a simpler model system in order to explore further this relation between the geometrical free volume and the particle DW factor.

## 2. The Model and Algorithm

In this paper we study a well-characterised model of a two-dimensional glass-forming alloy.[10] The equimolar mixture consists of particles interacting via a purely repulsive potential of the form $u_{ab}(r) = \varepsilon(\sigma_{ab}/r)^{12}$, where $\sigma_{12} = 1.2 \times \sigma_{11}$ and $\sigma_{22} = 1.2 \times \sigma_{11}$. All units quoted will be reduced so that $\sigma_{11} = \varepsilon = m = 1.0$ where $m$ is the mass of both types of particle. Specifically, the reduced unit of time is $\tau = \sigma_{11}(m/\varepsilon)^{\frac{1}{2}}$. The equations of motion of 1024 particles are propagated in the canonical ensemble using a generalised leapfrog implementation of the Nosé-Poincaré-Andersen Hamiltonian [11]. A square simulation cell with periodic boundary conditions is used, with the average pressure fixed at 13.5.

The particle Debye-Waller factors are calculated over an interval of $10\tau$ at $T = 0.4$. As mentioned above, this time corresponds to a point in the middle of the plateau region in the mean-squared displacement and in the incoherent scattering function. For reference, the structural relaxation time $\tau_e$ (the time at which the incoherent intermediate scattering function has decayed to $1/e$) is $1000\tau$ at $T = 0.4$.

Following Sastry et. al. [12], we define the free volume of a particle $i$ as the area accessible to that particle with all its neighbours fixed. For the purpose of this calculation we create a mapping to a hard-particle system by using a temperature dependent effective hard disc diameter corresponding to the distance of closest approach of two particles, as identified by the distance at which the respective pair distribution function first exceeds 0.01. At $T = 0.4$ this corresponds to a distance of $0.9\sigma_{ab}$ (where $ab$ stands for 11, 12 or 22, as appropriate). We note that the relative ordering of particles by free volume is fairly insensitive to small changes of this effective hard disc diameter.

We have calculated the distribution of free volume for each initial configuration and for the local potential energy minimum (the inherent structure) obtained when each initial configuration is used as the start of a conjugate gradient minimisation of the energy. The inherent structures appeared to produce a slightly greater degree of clustering of the free volume and so we shall only report the free volume for the inherent structures.

To account for the differences in particle size, we scale the free volume for each particle by $\pi.\sigma_{aa}^2/4$, where $a$ is the species of that particle. The scaled free volume $v_i$ has the added attraction that there exists a very local correlation between this geometrical measure and the potential energy $u_i$ of each particle $i$ in both the inherent and instantaneous structures. (Note that in calculating $u_i$ we consider only interactions with particles within a distance $g_{ab}$ from $i$, where $g_{ab}$ is the position of the first minimum in the respective pair distribution function. We find that this accounts for greater than 99.5% of the total potential energy.) In Figure 1 we plot the raw and scaled free volumes against the local potential energies of particles in the inherent structure. Unlike the raw free volumes, the scaled free volumes from the two species produces a single smooth curve when potted against the energy. This curve is well described by the expression $u_i = 1.1464 \, v_i^{-\frac{1}{2}}$. From here on we shall consider only the scaled free volume and shall refer to this simply as the free volume.

## 3. Results and Discussion

We have calculated the particle DW factor and free volume per particle for 10 configurations at $T = 0.4$. A run time of $75\tau_e$ separates each configuration. Shown in Figure 2, we find a smooth monotonic relation between the particle DW factor and the average free volume, when particles are divided into 20 subsets according to their DW factors. For values of the DW factor up to ~0.022 this relationship is linear, a result analogous to that found in ref. 9 for a 3D "bead-spring polymer" system. There appears to be an upper bound on the average particle free volume of about 0.06. This is visible as a plateau for DW factors above 0.03. We note that the maximum value of the DW factor, and hence the length of this plateau, increases with increasing time interval used to calculate the DW factor. These results certainly support the phenomenological results of a strong connection between free volume and the dynamics. The data also appear to support the idea that there is a well-defined threshold value of free volume or particle DW factor, above which large amplitude displacements occur.

This interpretation fails, however, when applied on a particle-by-particle basis. To see this, consider the standard deviation of the free volumes shown in the insert in Figure 2. Clearly, the free volume in a given dynamically defined subpopulation exhibits substantial fluctuations. Particles with wildly varying free volumes can exhibit similar values of the DW factor. We conclude that a particle's mobility, as characterised here by the DW factor, is not the result of its geometric free volume. As the amplitude of a particle's DW factor is a measure of the degree to which it is constrained by its surroundings, we conclude that the geometric free volume of that particle can only provide a haphazard glimpse of the degree of that constraint.

Our conclusion, that the variations between particles in terms of their geometric free volume cannot explain the variations observed in their DW factors, is supported by consideration of the spatial distribution of the two quantities. In Figure 3 we show contour maps for the free volume and the DW factor for a configuration at $T = 0.4$. There is a clear difference in the characteristic length scales of the distributions with the DW factors exhibiting significantly stronger clustering than the free volume.

We have quantified this observation by the following cluster analysis. To measure the spatial heterogeneity of a given property $A_i$ of particle $i$ we 'tag' the 10% of particles (102 particles in this case) with the largest values of $A$. We then assign each tagged particle $i$ (species $a$) to a cluster if it lies within a distance of $g_{ab}$ to any tagged particle $j$ (species $b$) already in that cluster, where $g_{ab}$ is the position of the first minimum in the corresponding pair distribution function. When all the tagged particles have been assigned to a cluster we count the number of clusters and calculate the variance in the number of particles per cluster. The maximum variance possible for a given number of clusters $N$ occurs when $N-1$ clusters consist of one particle and one cluster consists of $102 - (N-1)$ particles and is given by the relationship $\max(\sigma^2) = -205 - 10404/N^2 + 10608/N + N$. A random distribution without any spatial correlation will produce a large number of clusters with a corresponding small variance. A heterogeneous distribution will produce a smaller number of clusters.

In Figure 4 we have plotted the results of the cluster analysis for the free volume and the particle DW factor for the 10 configurations. Particles with high free volume show no significantly greater clustering than an equal number of randomly selected particles. In contrast particles with high DW factor show significantly more clustering. These results, in addition to highlighting the absence of any significant correlation between a particle's free volume and its DW factor, point to the source of the problem. The clear spatial clustering of those particles with large DW factors is evidence of the cooperative character of even this short time dynamics. The geometrical free volume is strictly a single particle property and, as such, fails to capture the subtle configurational features that result in enhanced local motion.

If the geometric free volume fails as a predictor of the local dynamic heterogeneity because the latter relies strongly on non-local correlations, can we improve the relevance of the free volume by using a suitable spatial average? For example, Qian et al [13] found that there was an optimal local averaging length (a coarse graining length) for which the Pearson's correlation coefficient of density and a residence time was maximised.

We have coarse grained the free volume and the local DW factor by assigning to each particle the value of the relevant property averaged over the local values for that particle and of the particles lying within a distance $r$ of that particle. We have considered values of $r$ in the range $0 \leq r \leq 10\sigma_{11}$. The degree of clustering as measured in Figure 2 increases steadily with increasing $r$. This is a trivial consequence of the coarse graining. The clustering observed in the particle DW factor is approximately reproduced in the coarse grained free volume for $r = 2$.

To measure correlation, we use Spearman's rank-order correlation coefficient $K$.[14] This calculates a linear correlation coefficient of ranks rather than values. For the case without coarse graining we find a value of $K = 0.40$, averaged over the 10 configurations studied. Readers are reminded that we have already demonstrated that there is no strong correlation between the scaled free volume and the particle DW factor through the comparison of spatial maps and the cluster analysis. As shown in Figure 5, we find no increase in the average correlation between the free volume and particle DW factors on coarse graining. The short comings of free volume as a predictor of dynamics, we conclude, are not to be corrected by simple spatial averaging.

## 4. Conclusions

In this 2D glass-former, having a larger free volume does not cause a particle to exhibit larger amplitude fluctuations in position. It does, however, increase the likelihood that the reduced local constraints necessary for large amplitude motion might apply. Even over the short time scales studied here, collective (i.e. non-local) processes are important and these are not well correlated with a purely local measure such as free volume. For this reason, we believe that the results reported here are likely to be common to many glass formers. In most of its popular usages, however, the phenomenological free volume refers, not to an explicit geometrical volume, but rather to a reduction of mechanical constraints on particle motion. In this sense, the particle Debye-Waller factor, defined in this paper, probably provides the better match since it is an explicit measure of particle constraint, even if it lacks a purely geometric definition. If one accepts this proposition, i.e. it is the particle DW factor rather than the geometrical free volume that provides the better microscopic expression of the phenomenological free volume, then the outstanding question for developing a microscopic treatment of dynamics in glassy materials is to see if there exists a method for predicting the particle Debye-Waller factors from a given configuration that is alogorithmically simpler than the dynamic averages presented in this paper.

## Acknowledgements

We would like to acknowledge the support of the Australian Research Council through the Discovery grant program.

Figure Captions

Figure 1. The relation between potential energy $u_i$ and (a) raw free volume (b) scaled free volume $v_i$ for particles in 10 configurations at $T = 0.4$. Scaling the raw free volumes by $\pi.\sigma_{aa}^{2}/4$, where $a$ is the species of each particle, collapses the data onto a single curve that is well described by the relation $u_i = 1.1464\ v_i^{-\frac{1}{2}}$.

Figure 2. Free volume as a function of particle Debye-Waller factor. Data for ten configurations at $T = 0.4$ have been pooled together, and the particles divided into 20 subsets according to their DW factors. Each subset is represented by a point in the graph. Error bars in the main graph represent one standard error. The inset shows the same data but with error bars corresponding to one standard deviation.

Figure 3. Contour plots of the spatial distribution of (a) free volume and (b) particle Debye-Waller factors for a configuration at $T = 0.4$. There is some spatial correlation between regions with low free volume and low DW factors but not in general between regions of high free volume and high DW factors.

Figure 4. Cluster measures of spatial heterogeneity for particles with Debye-Waller factors and free volumes in the top 10%. Data points are shown individually for ten configurations at $T = 0.4$. Statistics obtained using random values are shown for comparison. The dotted line represents the maximum variance possible for a given number of clusters (see text for more details).

Figure 5. Correlation between free volume and particle Debye-Waller factor as a function of coarse graining radius $r$. Correlation coefficients (Spearman's rank-order correlation) have been averaged over 10 configurations. The error bars represent one standard deviation.

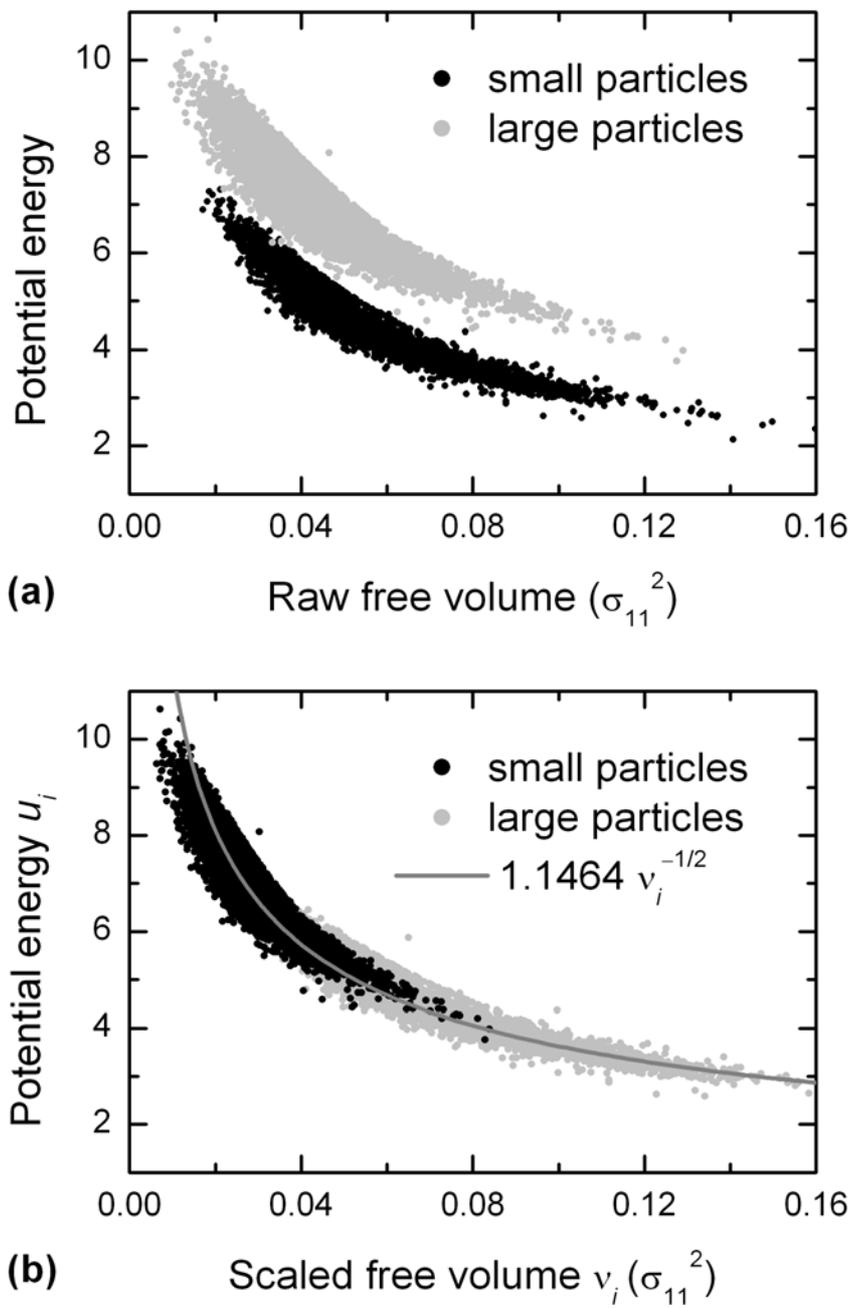

**(a)**

**(b)**

Figure 1

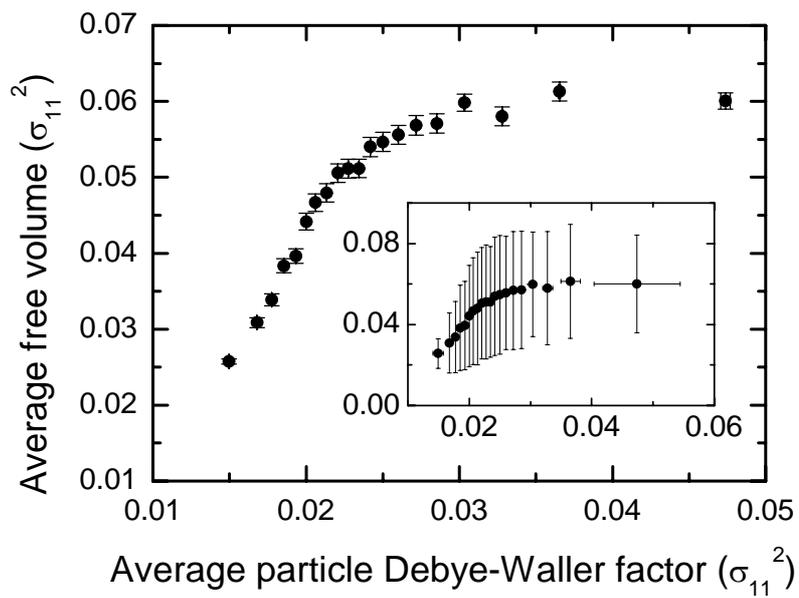

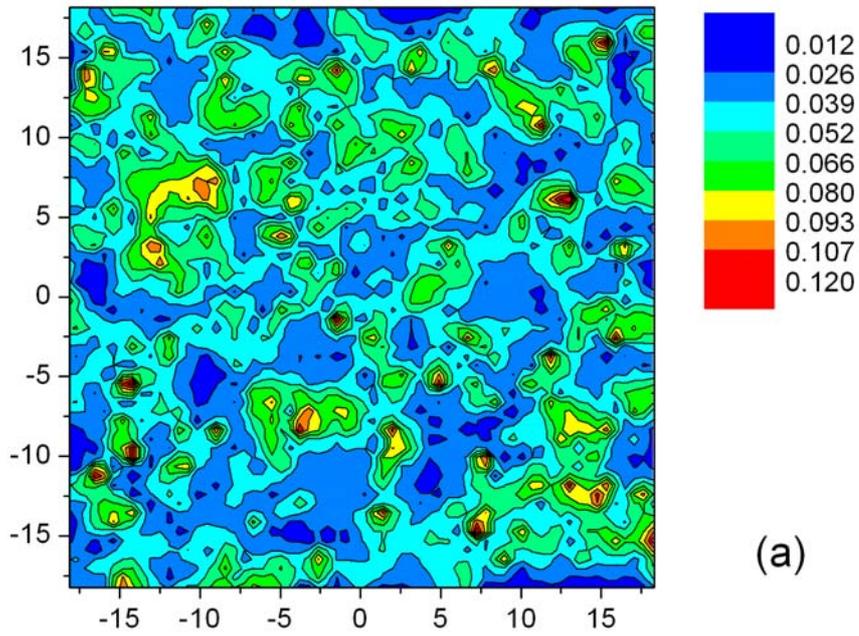

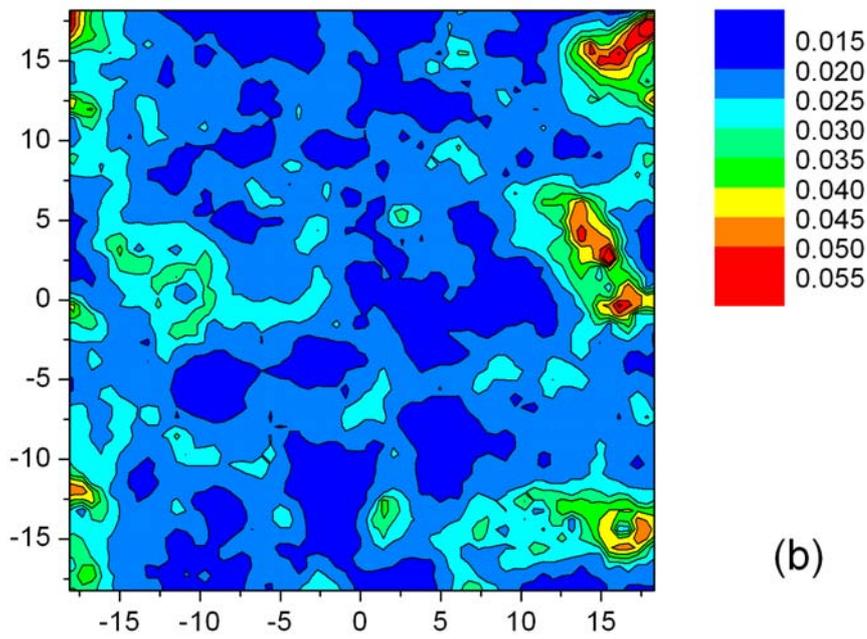

Figure 3

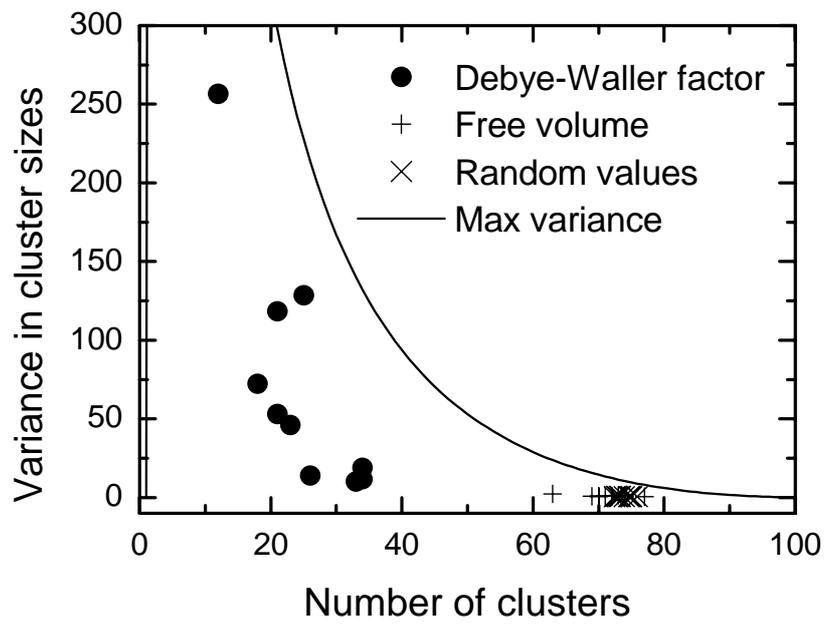

Figure 4

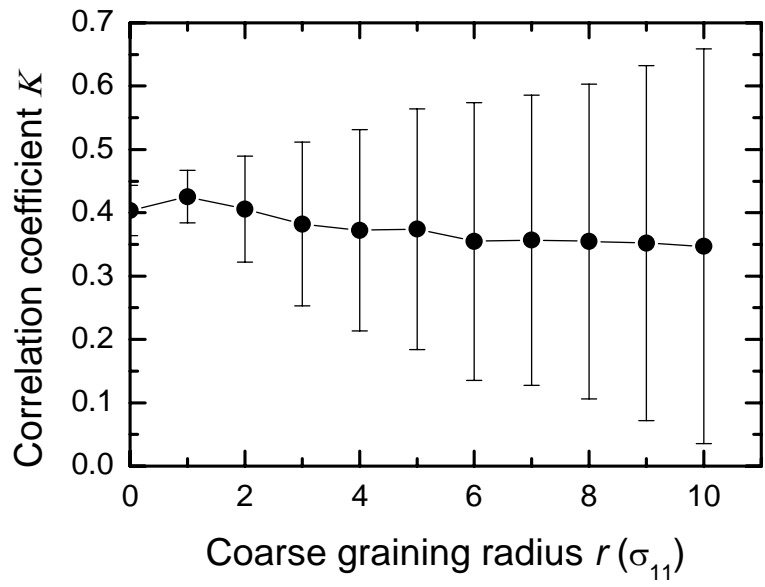

Figure 5